# Local optical chirality induced by near-field mode interference in achiral plasmonic metamolecules

*Andreas Horrer [†], Yinping Zhang [†], Davy Gérard [\*,†], Jérémie Béal[†], Mathieu Kociak [‡], Jérôme Plain[†] and Renaud Bachelot [†,§]*

[†] Light, Nanomaterials, Nanotechnologies (L2n), Institut Charles Delaunay, CNRS, Université de Technologie de Troyes, France

[‡] Laboratoire de Physique des Solides, Bâtiment 510, UMR CNRS 8502, Université Paris Sud, Orsay 91400, France

[§] Sino-European School of Technology, Shanghai University, 20444 Shanghai, China

[\*] email: davy.gerard@utt.fr

ABSTRACT. When circularly polarized light interacts with a nanostructure, the optical response depends on the geometry of the structure. If the nanostructure is chiral (i.e. it cannot be superimposed on its mirror image) then its optical response, both in near-field and far-field, depends on the handedness of the incident light. In contrast, achiral structures exhibit identical far-field responses for left- and right-circular polarization. Here, we show that a perfectly achiral nanostructure, a plasmonic metamolecule with trigonal $D_{3h}$ symmetry, exhibits a near-field response that is sensitive to the handedness of light. This effect stems from the near-field interference between the different plasmonic modes sustained by the plasmonic metamolecule under circularly polarized light excitation. The local chirality in a plasmonic trimer is then experimentally evidenced with nanoscale resolution using a molecular probe. Our experiments demonstrate that the optical near-field chirality can be imprinted into the photosensitive polymer, turning the *optical* chirality into a *geometrical* chirality that can be imaged using atomic force microscopy. These results are of interest for the field of polarization-sensitive photochemistry.

KEYWORDS. Chirality, plasmonic metamolecules, chiral plasmonics, photopolymers, circular polarization

Chirality, following Lord Kelvin's definition, refers to the property of an object that cannot be brought into congruence with its mirror image. The study of the interaction of light with chiral structures is now a two-century-long story, that can be traced back to seminal experiments in early 19th century by Arago, Biot and Pasteur. The latter evidenced how the two enantiomers of tartaric acid were acting on polarized light [1]. Actually, circularly polarized light can be thought of as *chiral light*, exhibiting two mirrored forms, i.e. the left-circular (LCP) and right-circular (RCP) polarizations. With the advent of nano-optics, artificial chiral nanostructures and metamaterials, such as helices [2], spirals [3], ramps [4], assemblies of nanoparticles (either colloidal [5] or lithographed [6]) and twisted stacks of nanorods arrays [7], have been developed. These structures exhibit artificial optical activity or circular dichroism that can be orders of magnitude higher than their natural counterparts and are tunable from the infrared to the visible region (see [8] for a recent review). Chiral nanostructures can hence be used as circular polarizers [9], phase plates [10] or to enhance chiroptical effects [11]. Of particular interest are two-dimensional (planar) chiral nanostructure. Strictly speaking, no 2D structure can be chiral, as it would be chiral only in a 2D universe. However, 2D-chiral structures do exhibit circular dichroism (i.e. their optical response depends on the handedness of the light) and they can be used to enhance chiroptic effects thanks to an effect known as super-chirality, or "chiral hot-spots" [12,13]. Planar chiral nanostructures can also exhibit chirality in their non-linear response [14]. These effects are prominent in metallic (plasmonic) nanostructures but can also appear in dielectric resonant nanostructures.

Optical chirality can also be observed in *achiral* structures [15,16,17]. In such structures, chirality can appear due to the symmetry breaking brought by oblique illumination, a tilt of the plane of the structure or off-axis polarization (extrinsic chirality) [18]. More surprisingly, optical chirality can be observed in achiral nanostructures even under symmetrical illumination. A remarkable illustration of this effect has been recently reported by Zu and co-workers first in V-shaped metal nanostructures and more recently in rectangular rods [19,20]. These resonators sustain two plasmonic eigenmodes overlapping in energy and whose interference under circularly polarized excitation induces an electromagnetic field distribution around the structure that depends on the handedness of the excitation (the near-field intensity distributions under LCP and RCP being mirror images). This effect has been called "hidden chirality" by Zu *et al.* to emphasize the fact that the chirality in their system was not of geometrical origin. A very similar interference effect was previously described by Abashal and coworkers, who demonstrated that an incoming linear polarization could be converted into circular polarization using plasmonic antennas sustaining two spectrally overlapping orthogonal modes [21].

In this Letter, we unveil optical chirality inside a highly symmetric plasmonic metamolecule (or plasmonic oligomer). The system consists of three gold nanodisks (or "meta-atoms") arranged as an equilateral triangle. The near-field coupling between each individual meta-atom leads to hybridized eigenmodes that can be described using group theory. At a specific wavelength, the phase difference between the excited modes reaches 90°, which once added to the ±90° phase-shift between the two linear components of circular polarization leads to handedness-dependent field distribution around the oligomer. We then provide experimental evidence of this local optical chirality with nanoscale imaging of the intensity distribution inside a plasmonic trimer. The imaging method is based on an azobenzene nanomotor molecular probe and yields ~20 nm resolution. A clear near-field contrast is experimentally evidenced between left- and right-circular polarizations.

**(a)**

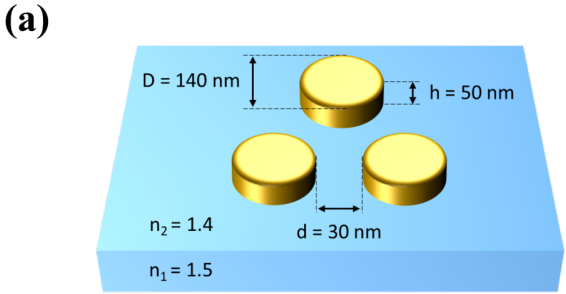

**(b)**

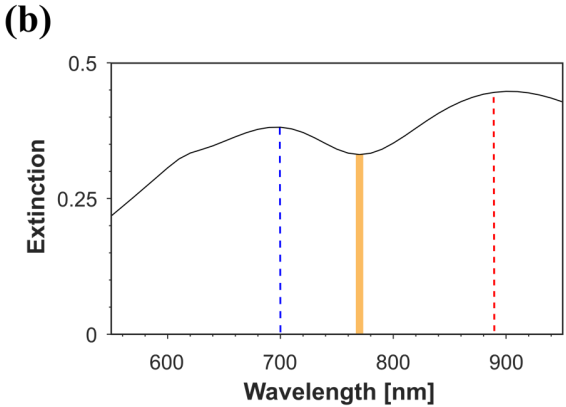

**Figure 1.** a) Schematic of the investigated plasmonic metamolecule. b) FDTD-simulated extinction spectrum of the trimer (the spectrum is identical independently of the incoming polarization). The blue and red vertical lines mark the position of the resonances. The yellow thick line corresponds to the position where the modes overlap and interfere ($\lambda$=770 nm).

We start by presenting simulations of the structure under investigation, which is depicted in **Figure 1a**. It consists of three identical gold nanodisks forming an equilateral triangle. The diameter of the disks is D = 140 nm, their height is 50 nm, they are separated by a gap distance $g$ = 30 nm and they are located on top of a glass substrate (refractive index $n_1$ = 1.5). We assume that the surrounding medium has an index of $n_2$ = 1.4, in order to take into account the experimental configuration we will use later to characterize the structures. The structure is obviously achiral and its far-field response is not sensitive to the handedness of the incident circular polarization. This is evidenced by the simulated extinction spectrum of **Figure 1b**, which is independent of the incident polarization. In contrast, the near-field response (i.e. the local electric field intensity) does exhibit a sensitivity to the incident polarization. **Figure 2** shows the computed near-field intensity maps for the different polarizations: linear horizontal (**Figure 2a**), linear vertical (**Figure 2b**), and both circular polarizations (**Figure 2c-d**). Intricate field distributions, associated with the hybridization of the plasmonic excitations borne by each nanoparticle, are observed. What is more surprising is the result for circularly polarized light, as the maps for LCP and RCP appear to be different. Although the field distribution inside the gap is the same for both signs of the circular polarization, the field in the outside part of the trimer clearly depends on the handedness of the light, the lobes of higher intensity around each nanoparticle being "twisted" in the direction of the polarization and following its sense of rotation. In contrast, the far-field spectrum is independent of the polarization of the excitation. We therefore call this phenomenon *local chirality*, to stress the fact that such chirality is confined to the near-field and cannot be observed in the far-field, unless a dedicated technique is used to probe the near-field.

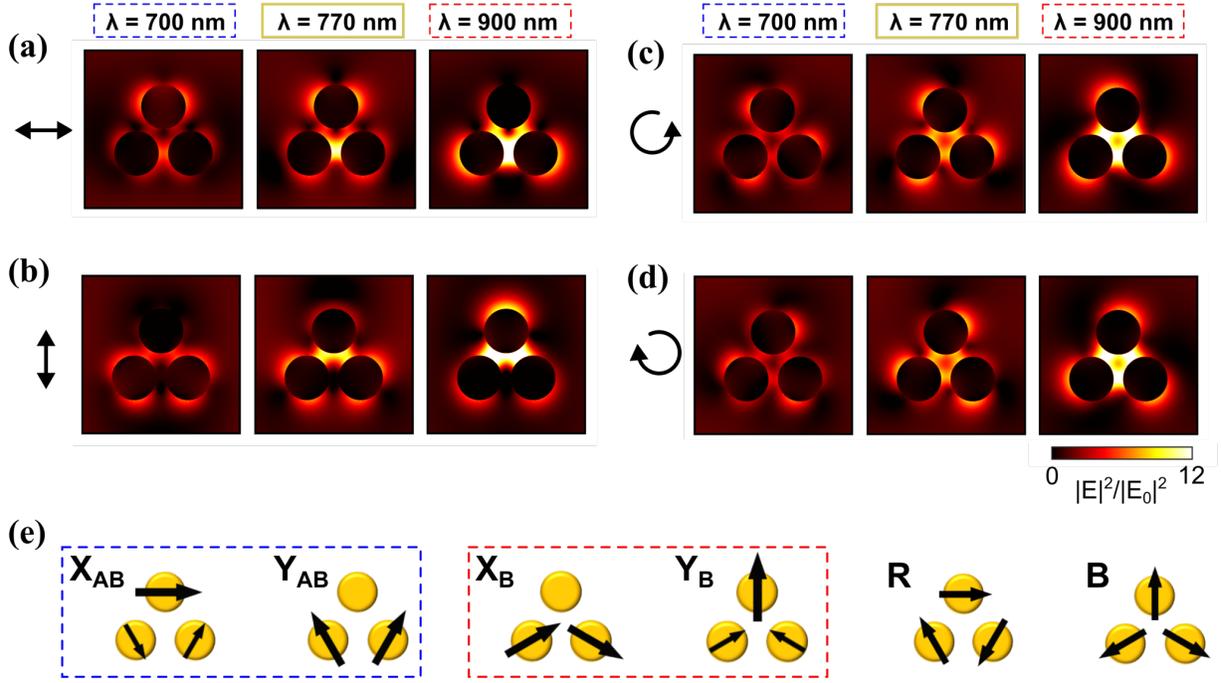

**Figure 2.** Near-field intensity maps for wavelengths of 700 nm, 770 nm, and 900 nm with an excitation with a) X-polarization, b) Y-polarization, c) left circular polarization, and d) right circular polarization. e) Dipole base modes of a $D_{3h}$ symmetric plasmonic trimer. The thickness of the arrows corresponds to the strength of the electric dipole. **$X_B$** (resp. **$X_{AB}$**) corresponds to the X-polarized bonding (resp. anti-bonding) mode (similar notations for Y-polarization). **R** is a rotation (or azimuthal) mode and **B** the breathing (or radial) mode.

In order to analyze these results, we need the eigenmodes of the plasmonic oligomer. These eigenmodes have been derived by Brandl *et al.* [22] using group theory. Neglecting the presence of the substrate, the plasmonic trimer pertains to the *$D_{3h}$* (ideal) point-group. The irreducible representations of the *$D_{3h}$* point-group can be used to obtain a set of symmetry adapted coordinates (SACs). This means that these SACs form an orthogonal basis for the electromagnetic excitations sustained by the trimer. The six SACs corresponding to *in-plane* dipole excitations of the meta-atoms are represented in **Figure 2e**. The arrows indicate the dipole moment carried by each nanoparticle. There are two modes with a total dipole moment in X-direction. In the limit of uncoupled discs, they are energetically degenerate and add up to a dipolar excitation in X-direction in each disc. For non-zero coupling strength the degeneration is lifted and the excitation at the lower energy is referred to as a bonding mode (**X$_B$**), while the mode at the higher energy is called an antibonding mode (**X$_{AB}$**). In the same way there are a bonding (**Y$_B$**) and an antibonding mode (**Y$_{AB}$**) with a total dipole moment in Y-direction. The rotation (**R**) and breathing (**B**) excitations exhibit zero total dipole moment and are therefore dark modes. This implies they will be weakly excited using linear or circular polarizations. Each electromagnetic excitation of the plasmonic trimer can hence be described using the four remaining SACs as an orthogonal basis. The peaks in the extinction spectrum can be attributed to the degenerate bonding and antibonding modes in X- and Y-directions. In the field maps in **Figure 2a-b** the SACs manifest themselves in the shape of the near-field, with the lower energy bonding modes being predominantly present at 900 nm and the antibonding modes at 700 nm.

Following Alegret *et al.* [23] it is possible to project any excitation of the plasmonic trimer onto the SAC basis by assuming that each particle bears a single dipole moment. This dipole moment can be retrieved from the FDTD calculations by using the following procedure: (i) Calculating the complex surface charge distribution (from the divergence of the electric near-field) in a plane located at mid-height of the meta-atoms ; and (ii) Vectorially adding all the charges in each disc to get their respective dipole moment vectors. Animated versions of these dipole moment vectors at 5 different wavelengths as a function of the phase are shown in the Supporting Information as **Movies 1-5**, for LCP and RCP excitation. Their projection on the SAC basis yields their relative contributions $c_i$ and phases $\varphi_i$. For this, the excitation state in the trimer is described by a vector $\boldsymbol{P}(\lambda)$ with six components, representing the complex dipole moments in x- and y-direction for each of the three particles. Solving the linear equation $\sum_n c_n(\lambda) e^{i\varphi_n(\lambda)} (\boldsymbol{SAC}_n) = \boldsymbol{P}(\lambda)$, where $\boldsymbol{SAC}_n$ is the 6-dimensional vector associated with the base mode $n$, gives the coefficients as a function of the wavelength. **Figure 3** shows the case for right circularly polarized excitation. Here all four dipolar SACs contribute to the total particle polarization. The relative weights (**Figure 3a**) as well as the relative phases (**Figure 3b**) of the SACs can be seen. Please note that we also computed the weight of the dark modes **R** and **B**: their relative weight was found to be much weaker than the bright modes (see **Figure S1**). Therefore, in the following we only consider the bright modes. The two degenerate antibonding modes **X**$_{AB}$ and **Y**$_{AB}$ have a stronger contribution at shorter wavelengths, while the two degenerate bonding modes **X**$_B$ and **Y**$_B$ are dominant for longer wavelengths. Under circularly polarized excitation, there is a constant phase difference of ±π/2 between the perpendicular modes in X- and Y-direction (for instance between **X**$_B$ and **Y**$_B$). Moreover, due to the different resonance wavelengths of the bonding and antibonding modes, there is a certain wavelength range (corresponding to the dip in the extinction spectrum in **Figure 1b**) where they have different relative phases (for instance between **X**$_B$ and **X**$_{AB}$). This phase difference depends on the coupling strength between the individual dipoles in the metamolecule. For the trimer with the aforementioned dimensions, a maximum phase difference of about π/2 at a wavelength of about 770 nm is observed. For perpendicular bonding and antibonding modes this means that the total phase differences add up to about π between **X**$_B$ and **Y**$_{AB}$ and to about 0 between **Y**$_B$ and **X**$_{AB}$ for RCP (**Figure 3b**). In the other case, for LCP, the phase differences add up to about 0 between **X**$_B$ and **Y**$_{AB}$ and to about π between **Y**$_B$ and **X**$_{AB}$. This leads to mirrored near-field profiles for LCP and RCP. For wavelengths above or below 770 nm the total phase differences deviate from this condition. This leads to a less distinct near-field difference between LCP and RCP excitation. To go further, we show in **Figure 3c** the near-field profiles and in **Figure 3d** the *complex* dipole moment vectors for each particle in the trimer at different wavelengths. The dipole moments are depicted in a way proposed in [24]. The ellipses are the curves traced by the tip of the dipole moment vector over one period, the phase being color-coded. An animated version of the images can be found as **Movie 6** in the **Supporting Information**. The relative contributions and phase relations of the modes lead to different shapes of the ellipses at different wavelengths. These polarization ellipses give the polarization state of each meta-atom of the plasmonic metamolecule. In **Figure 3**, five different situations are shown. Starting from the lowest wavelength (wavelength 1, $\lambda = 600$ nm), each particle exhibits an almost circularly-polarized response. In this situation there is no phase difference between the antibonding and bonding modes and the contributions of each mode are nearly equal. There is however a phase difference of $\pm\pi/2$ between the perpendicular modes in X- and Y-direction, which comes from the circular polarized excitation. This leads to dipole moments that rotate in phase along a circle in each disc (see also **Movie 1**). As a consequence, the near-field profile is nearly equally distributed around the discs in the trimer and the field maps look similar for RCP and LCP excitation. When the wavelength increases, the polarization state becomes more and more elliptical. At a wavelength of $\lambda = 700$ nm (marked as (2) in the figure) the contributions of the two antibonding modes in X- and Y-direction are dominant. Furthermore, there is a phase difference between the bonding and antibonding modes which is in the range of $0 < \Delta\varphi < \pi/2$. Consequently, the tip of each individual dipole moment is following an ellipse over one period in each disc. It should be noted that the phase for the situation where the dipole moments are aligned with the major axis of the ellipse is shifted by $\pm 2/3\pi$ for the three discs. The *total* dipole moment in the trimer is therefore following the rotation sense of the exciting circular polarized light (see **Movie 2**).

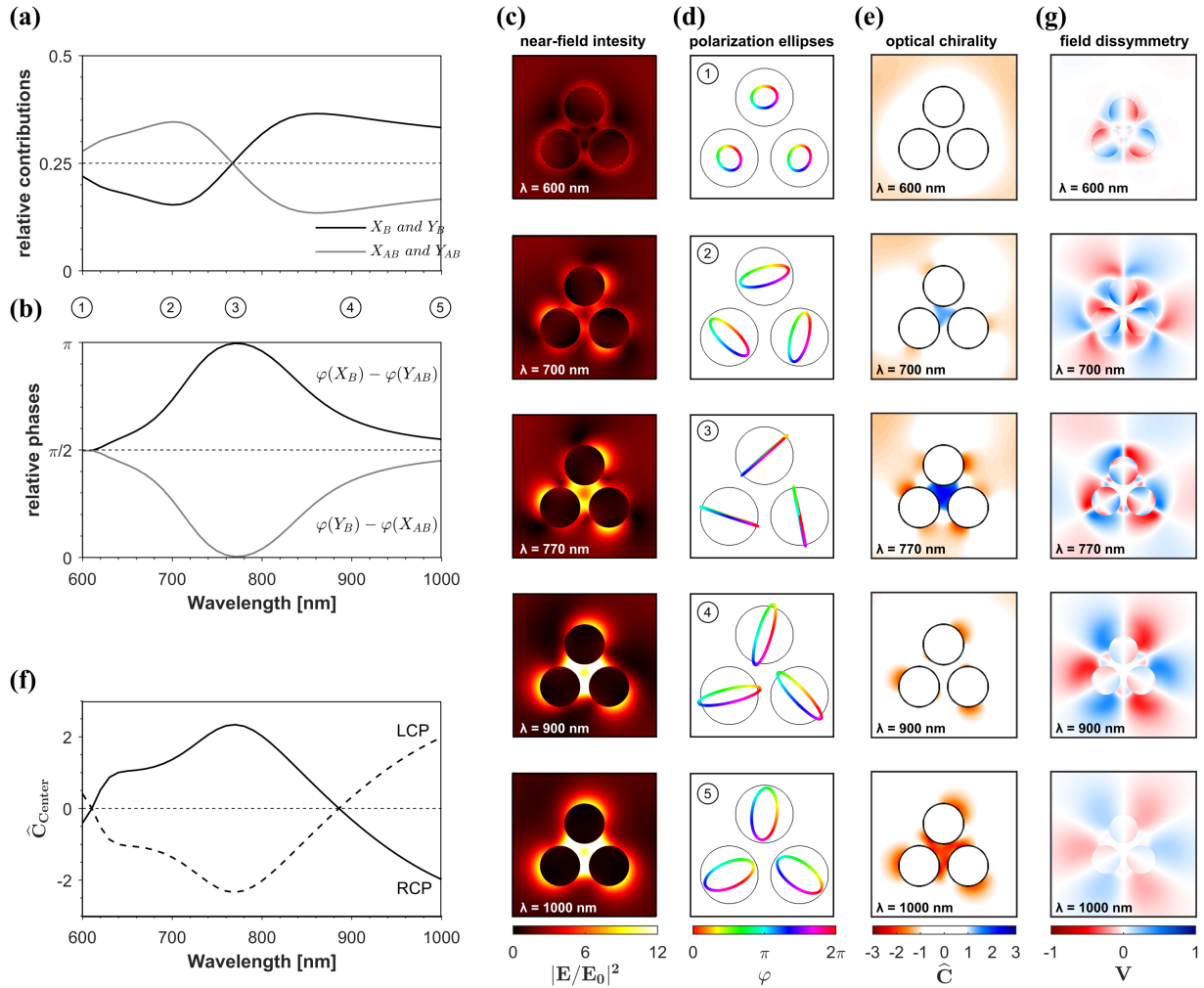

**Figure 3.** (a) Relative contributions of the bonding modes $X_B$ and $Y_B$ (black solid line) and the antibonding modes $X_{AB}$ and $Y_{AB}$ (light grey line) for RCP excitation. (b) Relative phases of the perpendicular bonding and antibonding modes $(X_B) - (Y_{AB})$ in black and $\varphi(Y_B) - \varphi(X_{AB})$ in light grey for RCP excitation. (c) Maps of the electric near-field intensity, (d) polarization ellipses and (e) maps of the normalized optical chirality density $\hat{C}$ for RCP excitation at five different wavelengths. (f) Optical chirality density $\hat{C}$ computed at the center of the metamolecule under LCP and RCP excitation. (g) Maps of the electric field dissymmetry factor $V$ computed at five different wavelengths.

When λ=770 nm (wavelength 3), we can see a linear excitation in each disc (see also **Movie 3**). This is the result of all modes contributing equally and a phase difference of π/2 between the bonding and the antibonding modes. Together with the π/2 phase difference from the circular polarized excitation this adds up to a total phase difference of π between **X$_B$** and **Y$_{AB}$** and 0 between **Y$_B$** and **X$_{AB}$**, as described above. For LCP excitation it is vice versa (0 between **X$_B$** and **Y$_{AB}$** and π between **Y$_B$** and **X$_{AB}$**), leading to a linear excitation in each disc, which is perpendicular to the one with RCP excitation. For longer wavelengths (wavelengths (4) and (5) in the figure), the contributions of the bonding modes are dominant and the polarization state of each meta-atom becomes elliptical again. Note that the polarization ellipses under excitation with X, Y and LCP polarizations for the same wavelengths can be found in **Figure S2**.

Then, we investigated the strength of the local chirality inside the oligomers. The optical chirality (or optical chirality density) $C$ is the most common quantity used to quantify near-field chirality as it is related to the rate of excitation of chiral molecules [12,25]. It is defined as $C = -\frac{\omega\epsilon_0}{2} Im(\boldsymbol{E}^* \cdot \boldsymbol{B})$. Following Schäferling and coworkers, we subsequently normalized this value using the corresponding value of a simulation without structure in order to obtain the optical chirality enhancement $\hat{C}^{\pm} = C^{\pm}/|C_0^{\pm}|$, where the + and − signs refer to LCP and RCP, respectively and $C_0$ is the value of the optical chirality without the metamolecule. Regions where $|\hat{C}| > 1$ correspond to the so-called superchiral fields. **Figure 3e** shows maps of $\hat{C}$ for metamolecules with a gap size of 30 nm. The colormap has been chosen so that regions exhibiting superchirality appear in color. We also plotted the value of $\hat{C}$ in the center of the structure as a function of the wavelength in **Figure 3f**. It is clear from these figures that we do not observe superchiral fields at short wavelengths where each meta-atom bears a circular polarization. Superchirality increases with the wavelength and reaches a maximum at 770 nm, where we also observe a clear contrast in the sign of $\hat{C}$ between the center of the structure and the sides of the metamolecule. At this wavelength, the optical chirality in the gap is opposite to the chirality of the exciting wave (i.e. the chirality is left-handed in the gap for RCP excitation). It is also interesting to note the change of sign of the value of $\hat{C}$ at the center. For wavelengths below ~900 nm $\hat{C}$ is negative. This means the field is twisted in a direction opposite to the excitation polarization. On contrary for longer wavelengths, the field is twisted in the same direction as the excitation. It can be related to the change of the dominant modes inside the structure. When the dipoles in each meta-atom are oriented mostly azimuthal, the chirality density in the center is positive, while when the dipoles are mostly radial, it is negative. Interestingly, as the local chirality arises from an interference-based effect, its strength (i.e. the value of the chirality density) can be tuned by controlling the interparticle distance (see **Figure S3**).

The optical chirality $\hat{C}$ is proportional to both electric and magnetic fields. In order to quantify the chirality of the electric field only, it is possible to introduce a near-field contrast (or visibility) $V$ between RCP and LCP excitations. We define it similarly to Ref. [19] as the normalized difference of the field intensities for LCP and RCP:

$$V = \frac{|E_{LCP}|^2 - |E_{RCP}|^2}{|E_{LCP}|^2 + |E_{RCP}|^2}$$

The $V$ factor can be thought of as the local dissymmetry factor (*g*-factor) of the electric field intensity or, in other words, as the circular dichroism of the near-field intensity. Computed maps of $V$ are shown in **Figure 3g** for the same five wavelengths as before. The electric field intensity

*g*-factor reaches a maximum value of about 55% at λ = 770 nm (near the meta-atoms) while it is only ~15% at λ = 600 nm and at λ = 1000 nm. The *V* factor can quantify the chirality of interactions that depend mainly on the electric field intensity.

In order to experimentally demonstrate our numerical findings, we fabricated gold metamolecules on a glass substrate using electron-beam lithography. We obtained well-defined nanostructures with reproducible sizes and gap distances (see the SEM image in **Figure 4a**). Previous reports on the experimental characterization of the optical near-field chirality relied on near-field optical microscopy (SNOM) [17,26,27] and cathodoluminescence [20]. Here, we use a photosensitive azobenzene-containing polymer (PAP) to unveil the spatial distribution of the optical intensity around the plasmonic oligomer. Azobenzene molecules, such as disperse-red one (DR1), act as molecular nanomotors under illumination: the absorption of a photon by the molecule yields a change of conformation that induces a "worm-like" movement along the direction defined by its transition moment. If the azobenzene is grafted to a polymer backbone (such as PMMA), then this molecular movement leads to a large-scale displacement of matter, which can be tracked using atomic force microscopy (AFM). Previous studies have demonstrated that the lateral resolution of this technique is about 20 nm [28] and can map plasmonic hot spots [29]. Furthermore, PAP like PMMA-DR1 are sensitive to the polarization state [30] and to the orbital angular momentum carried by vortex beams [31]. A limitation of this characterization technique lies in the fact that the PAP film must be excited at a wavelength located inside the absorption band of the DR1 molecule (in the green part of the visible spectrum, see **Figure S4**, blue curve). A way to circumvent this issue is to use two-photon excitation. We recently demonstrated that under two-photon excitation, PAP systems have the ability to map hot-spots inside plasmonic oligomers resonating in the near-infrared [32]. The two-photon absorption spectrum of a PMMA-DR1 film is shown in **Figure S4** (red curve). In the following, we apply this nanoscale characterization technique to image the near-field chirality inside plasmonic metamolecules. Following the method described in [32], the metamolecules were covered by a 45-nm-thick PMMA-DR1 film. An AFM image (see Methods) of the PAP-covered nanostructures is shown in **Figure 4b**, as a reference of the initial situation. An exemplary far-field reflection spectrum (normalized to the reflection of the bare sample surface) recorded on an ensemble of PMMA-covered metamolecules is shown in **Figure 4c**. The peaks associated with the excitation of the bonding and antibonding modes can be clearly distinguished. A good agreement is found with the above simulated spectrum (**Figure 1b**). Then, we illuminated the metamolecules using a weakly focused laser beam from a Ti:Sa laser at λ=790 nm. This wavelength lies between the two peaks in the reflection spectrum in **Figure 4c**, corresponding to the position where the local chirality is expected to be the highest. Note that this wavelength does not match the maximum of two-photon absorption in the PAP film [32], but the relatively low absorption coefficient observed at 790 nm is still sufficient to trigger matter displacement, as it will be evidenced below. The laser beam is polarized either linearly or circularly (using a linear polarizer and a quarter wave-plate). The quality of the generated circular polarization has been carefully checked using an analyzer.

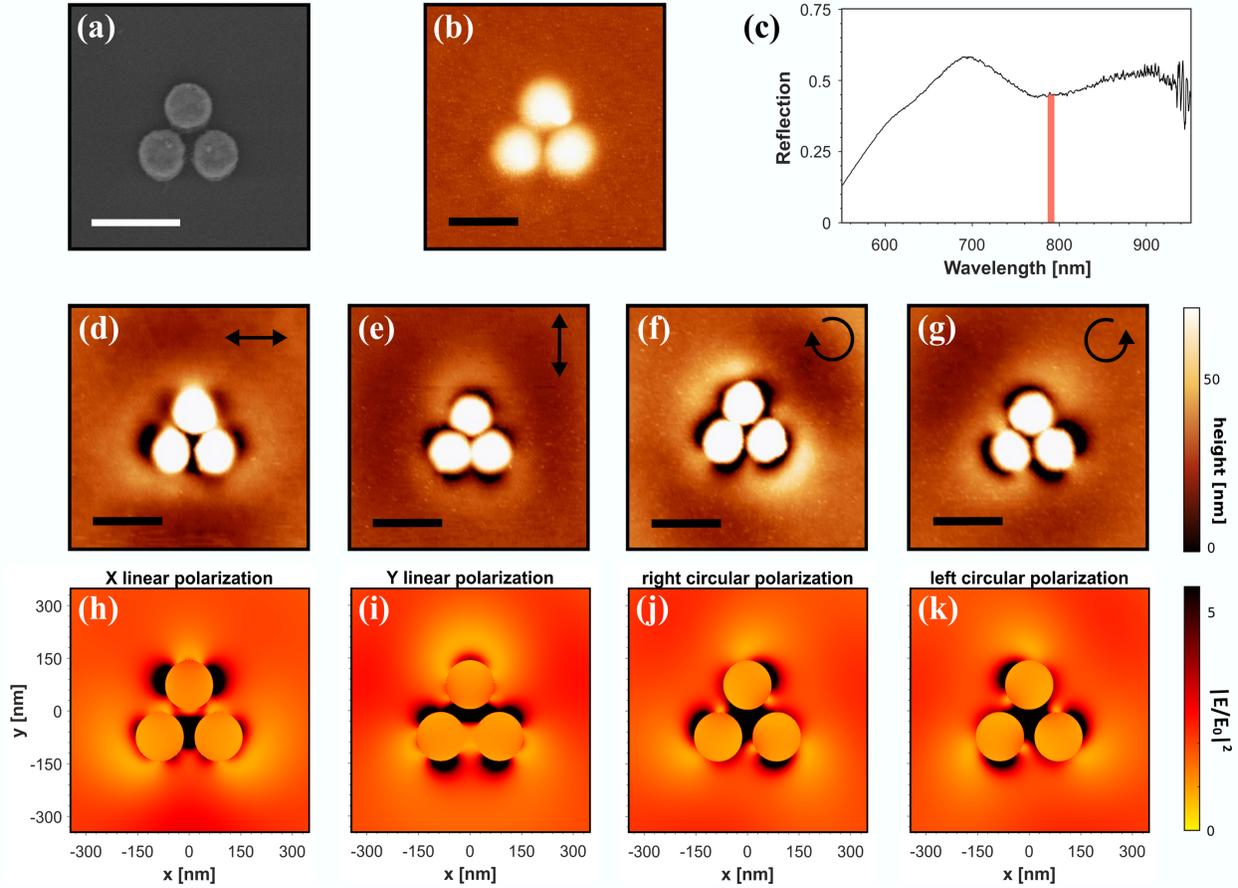

**Figure 4.** Experimental results: (a) SEM image of a gold metamolecule. (b) AFM image of the metamolecule after PAP film deposition. (c) Exemplary reflection spectrum recorded on an assembly of metamolecules. The thick red line marks the wavelength of the laser. (d-g) AFM images of the plasmonic metamolecules after illumination using (d) X-polarized excitation ; (e) Y-polarized excitation ; (f) RCP excitation and (g) LCP excitation. The scale bars on all images are 300 nm. (h-i) simulated results of the modulus of the component of the electric field in the direction of the corresponding excitation polarization. (h) $|E_x|^2$ with X-polarized excitation ; (i) $|E_y|^2$ with Y-polarized excitation ; (j) $|E_-|^2$ with RCP excitation; (k) $|E_+|^2$ with LCP excitation.

The results for four different incident polarizations (linear horizontal, linear vertical, LCP and RCP) are shown in **Figure 4d-g**, respectively. The images are the AFM topographical images recorded after illumination with polarized light. Dark areas correspond to regions where matter has escaped, indicating a higher in-plane local field intensity. Let us first focus on the linear polarization excitation (Figure 4d-e). It should be noted that we do not observe matter displacement only in the direction of the incident linear polarization, as one would expect if aligned dipoles were excited in each nanoparticle inside the oligomer. In contrast, we observe a field map which is linked to the dipole orientations of the SACs (due to the near-field coupling between the meta-atoms). It was shown in [32] that the imprinted profile has the shape of the component of the electric field oriented in the same direction as the excitation polarization. Therefore, below the experimental results we show the simulated field maps of $|E_x|$ for X-polarization and $|E_y|$ for Y-polarization. A very good agreement is found with the AFM topography, in particular the depleted regions in the AFM images (black regions) correspond well with the areas of high electric field, while the elevations of the resist (in yellow) can be identified as the areas with relatively low electric field. The image for Y-linear polarization (**Figure 4e**) exhibits a very good agreement with the FDTD calculation of the electric field. A slightly less nice agreement is found for the X-polarized case, which could be explained by small asymmetry in the fabricated structure (see below).

Now we turn our attention towards the case of circularly polarized excitation. The images for RCP and LCP excitation of the oligomer are presented in **Figures 4e** and **4f**, respectively. They appear to be different, showing that the PAP-covered metamolecule is sensitive to the handedness of the incident light. A closer look at the figure shows that the two images are almost mirrored, in agreement with the computed near-field intensity maps presented in **Figure 2**. It should be emphasized that a bare PAP film is not sensitive to the light handedness [33]. This sensitivity stems from the achiral oligomer, and the PAP film turns the near-field chirality into surface topography. After illumination with chiral light, the achiral structure hence becomes geometrically chiral due to the matter displacement. Interestingly, illuminating two initially identical oligomers with either LCP or RCP leads to two structures that are mirror images of each other, i.e. enantiomers. Following the same method as for linear polarization, we computed the components of the electric field in the circular polarization basis. We hence calculated the component of the electric field with LCP and RCP orientation using $E_\pm = E_x + E_y \cdot exp\left(\pm i\frac{\pi}{2}\right)$. Again, a nice agreement with the experiment is found, with slight discrepancies. The difference between the experimental and simulated near-fields can be accounted to slight asymmetry of the shape of the discs and their positions in the trimer. In such a case, the orientation of the dipoles in the base modes would change [34], due to the lifted degeneracy of the X- and Y-modes and a larger contribution of the rotation and breathing modes. This would lead to a different field distribution. Another possible source of asymmetry is the incident excitation, especially an imperfect (slightly elliptical) polarization state and/or a weak angle of incidence. As the near-field chirality stems for an interference effect, it is expected to be highly sensitive to any change in the excitation state. Let us also emphasize that the same behavior was observed over several different fabricated oligomers and the results are reproducible (see **Figure S5**). Altogether, our results confirm that the PAP film is sensitive to the electric field, in agreement with previous reports [28-32]. While the photochemical imaging system is not able to image the optical chirality *C*, it is sensitive to the near-field dissymmetry.

To conclude, we have shown how the near-field interference between the modes supported by a plasmonic metamolecule can create local optical chirality. This chirality stems for the near-field coupling between the individual meta-atoms. This phenomenon has then been experimentally measured using a photosensitive polymer as a molecular probe of the optical chirality. The experiments unveiled a clear dependence of the near-field response of the metamolecule upon the handedness of the incident light, leading to a clear near-field dissymmetry. Furthermore, this near-field dissymmetry is *imprinted* into the polymer film, efficiently transforming the optical chirality into geometrical chirality – the initially achiral structure becoming chiral after interaction with chiral light. Besides their potential to image and analyze the local chirality with a resolution on the nanometer scale, our results are of interest for the field of chiral photochemistry, including nanofabrication [36] and hot electron generation [37] driven by circularly polarized light.

METHODS

*Numerical simulations.* All simulations were performed using a commercial solution based on the finite difference time-domain method (FDTD Solutions from Lumerical Solutions).

*AFM imaging.* AFM images were recorded using a Veeco Bioscope II AFM microscope working in tapping mode and equipped with sharp tips (Nanosensors PPP-NCLR probe, with tip radius < 10 nm). A typical image size was $5 \times 5$ μm$^2$ and the scan rate was set to 0.5 Hz.

*Photosensitive azobenzene-containing polymer film illumination.* The PMMA-DR1 film was excited using a femtosecond Ti:Sa laser emitting at λ=790 nm (pulse duration ~200 fs). The beam was polarized using a linear polarizer, completed by a quarter-wave plate oriented at ±45° of the polarization axis to create circular polarization. Before impinging the sample, the laser beam was slightly focused using a 10x, NA=0.3 objective lens. The average power density was 700 mW/cm$^2$ and the exposure time was 60 s.

## ASSOCIATED CONTENT

The following files are available free of charge.

Supplementary information: Figures S1, S2, S3, S4 and S5 (PDF)

Movies 1-5 : movies showing the individual dipole moments of each meta-atom (thick arrows) and the total dipole moment of the metamolecule (thin arrow) vs. the phase for five different wavelengths: λ=600nm (Movie 1) ; λ=700nm (Movie 2) ; λ=770nm (Movie 3) ; λ=900nm (Movie 4) ; λ=1000nm (Movie 5). (avi)

Movie 6 : movie showing the polarization ellipses inside the meta-atoms vs. the wavelength (avi)

## AUTHOR INFORMATION

**Corresponding Author**

*davy.gerard@utt.fr

**Author Contributions**

AH performed all numerical simulations. YZ performed all experiments. AH, DG and MK proposed the near-field mode interference model. DG and AH wrote the manuscript. JB fabricated the sample. JP and RB defined and supervised the project. All authors discussed the results and contributed to the manuscript.

**Funding Sources**

Agence Nationale de la Recherche, under grants ANR-17-CE24-0039 (2D-CHIRAL), ANR-13-BS10-0013-03 (NATO), ANR-15-CE24-0036-01 (ACTIVE-NANOPHOT).

ACKNOWLEDGMENT

We would like to thank the anonymous reviewers of our manuscript for their valuable comments and suggestions. We also thank Frédéric Laux for his help on FDTD simulations. DG and MK are most grateful to Prof. Brouilly for his insights about chirality. YZ was supported by a PhD grant from the Chinese Scholarship Council. The plasmonic metamolecules were fabricated in the Nano'mat Platform ([www.nanomat.eu](www.nanomat.eu)), which is supported by the *Ministère de l'enseignement supérieur et de la Recherche*, the *Région Grand Est*, the *Conseil Général de l'Aube* and FEDER funds from the European Community. This work was also partially supported by the HPC Center of Champagne-Ardenne ROMEO.

ABBREVIATIONS
DR1, disperse red 1; PMMA, Poly(methyl methacrylate); RCP, right-handed circular polarization; LCP, left-handed circular polarization.